\documentclass[a4paper,twoside]{article}

\usepackage[T1]{fontenc}
\usepackage[utf8x]{inputenc}
\usepackage[english]{babel}
\usepackage[colorlinks=true, allcolors=blue]{hyperref}

\usepackage{listings}
\usepackage{xcolor}
\usepackage{authblk}
\usepackage[a4paper,top=3cm,bottom=2cm,left=3cm,right=3cm,marginparwidth=1.75cm]{geometry}
\AtBeginDocument{}

\usepackage{epsfig}
\usepackage{subcaption}
\usepackage{calc}
\usepackage{amssymb}
\usepackage{amstext}
\usepackage{amsmath}
\usepackage{amsthm}
\usepackage{multicol}
\usepackage{pslatex}
\usepackage{apalike}
\usepackage{algorithm2e}
\usepackage[bottom]{footmisc}
\usepackage{orcidlink}

\begin{document}

\definecolor{mGreen}{rgb}{0,0.6,0}
\definecolor{mGray}{rgb}{0.5,0.5,0.5}
\definecolor{mPurple}{rgb}{0.58,0,0.82}
\definecolor{backgroundColour}{rgb}{0.95,0.95,0.92}

\setlength{\fboxsep}{1pt}

\lstdefinestyle{CStyle}{
    backgroundcolor=\color{backgroundColour},   
    commentstyle=\color{mGreen},
    keywordstyle=\color{magenta},
    numberstyle=\tiny\color{mGray},
    stringstyle=\color{mPurple},
    basicstyle=\scriptsize\ttfamily,
    breakatwhitespace=false,         
    breaklines=true,                 
    captionpos=b,                    
    keepspaces=true,                 
    numbers=left,                    
    numbersep=5pt,                  
    showspaces=false,                
    showstringspaces=false,
    showtabs=false,                  
    tabsize=2,
    language=C,
    escapeinside={(*@}{@*)}
}

\title{Simple Power Analysis of Polynomial Multiplication in HQC}


\author{Pavel Velek\, \orcidlink{0009-0002-6069-9530}}
\author{Tomáš Rabas\, \orcidlink{0000-0002-0924-359X}}
\author{Jiří Buček\, \orcidlink{0000-0003-1359-4285}}
\affil{\{velekpav, tomas.rabas, jiri.bucek\}@fit.cvut.cz}
\affil{Faculty of Information Technology, Czech Technical University in Prague, Thákurova 9, Prague, Czech Republic}

\date{\small This manuscript was sent on November 11, 2025 for review to the conference ICISSP 2026, 12th International Conference on Information Systems Security and Privacy, \url{https://icissp.scitevents.org}}





\maketitle

\begin{abstract}
The Hamming Quasi-Cyclic (HQC) cryptosystem was selected for standardization in the fourth round of the NIST Post-Quantum Cryptography (PQC) standardization project. The goal of the PQC project is to standardize one or more quantum-resistant public-key cryptographic algorithms. In this paper, we present a single-trace Simple Power Analysis (SPA) attack against HQC that exploits power consumption leakage that occurs during polynomial multiplication performed at the beginning of HQC decryption. Using the ChipWhisperer-Lite board, we perform and evaluate the attack, achieving a 99.69\% success rate over 10\,000 attack attempts. We also propose various countermeasures against the attack and evaluate their time complexity.
\end{abstract}


\section{\uppercase{Introduction}}

\label{sec:introduction}

With the increasing threat posed by quantum computers, many new cryptographic schemes have been developed to provide security not only against attacks carried out using conventional computers but also against attacks performed with quantum computers capable of breaking widely used cryptographic algorithms. In response to this threat, the National Institute of Standards and Technology (NIST) started the Post-Quantum Cryptography (PQC) standardization project in 2016, with the aim of evaluating and standardizing one or more quantum-resistant public-key cryptographic algorithms. As part of the fourth round of this project, the Hamming quasi-cyclic (HQC) cryptosystem was selected for standardization on March 11, 2025.

In this paper, we present a single-trace simple power analysis (SPA) attack against HQC. The aim of our attack is to recover the private key by analyzing power consumption during polynomial multiplication that is performed at the beginning of HQC decryption. The target implementation of our attack is  the \emph{Additional implementation} of HQC included in the NIST Round~4 submission~\cite{HQCsubmission}.

The target implementation is particularly relevant as it is included in the PQClean library~\cite{pqclean}, which is a collection of clean implementations of post-quantum cryptographic algorithms, and is also in the liboqs library~\cite{liboqs}, a C library developed as part of the Open Quantum Safe project that aims to develop and integrate quantum-safe cryptography into applications.

The first published paper proposing a power side-channel attack on HQC is the paper \emph{A Power Side-Channel Attack on the CCA2-Secure HQC KEM}~\cite{10.1007/978-3-030-68487-7_8}. The attack presented in the paper uses a power side-channel to build an oracle with less than 10\,000 measurements. This oracle reveals whether the BCH decoder in HQC’s decryption algorithm corrects an error for a chosen ciphertext, allowing the recovery of a large part of the secret key.

The paper \emph{A Power Side-Channel Attack on the Reed-Muller Reed-Solomon Version of the HQC Cryptosystem}~\cite{10.1007/978-3-031-17234-2_16} presents an attack against the Reed-Muller Reed-Solomon (RMRS) version of HQC from the third round of the NIST PQC standardization competition. The authors of this paper, who also published  the previous paper~\cite{10.1007/978-3-030-68487-7_8}, adapt their earlier power side-channel attack to target the RMRS version of HQC.

Another attack, proposed in the paper \emph{A New Key Recovery Side-Channel Attack on HQC with Chosen Ciphertext}~\cite{10.1007/978-3-031-17234-2_17} also targets the RMRS version of HQC. In this paper, it is shown that it is possible to retrieve a static secret key of HQC by targeting the Reed-Muller (RM) decoding step with an electromagnetic side-channel attack.

A more recent paper, \emph{OT-PCA: New Key-Recovery Plaintext-Checking Oracle Based Side-Channel Attacks on HQC with Offline Templates}~\cite{cryptoeprint:2024/1715}, introduces an attack that first builds offline templates (OTs), and then queries a plaintext-checking (PC) oracle and uses the oracle’s responses together with the offline templates to recover secret information.

The first Soft Analytical Side-Channel Attack (SASCA) on HQC is presented in the paper \emph{Single trace HQC shared key recovery with SASCA}~\cite{Goy_Maillard_Gaborit_Loiseau_2024}. The attack exploits the polynomial multiplication performed in the Reed–Solomon (RS) decoder and can recover the shared secret key using only a single power trace. In comparison, our attack focuses on the polynomial multiplication that occurs before the RS decoder, which involves the private key and the ciphertext. The key difference is that our attack aims to recover the private key itself rather than the shared secret key.

The attack proposed in the paper \emph{Profiling Side-Channel Attack on HQC Polynomial Multiplication Using Machine Learning Methods}~\cite{bworldUpdate} targets the same implementation and polynomial multiplication as our work. However, the attack proposed in our paper does not rely on a profiling phase or any machine learning methods, while still achieving a similar high success rate. As a result, our attack presents a greater practical risk to real-world deployments.

Other recently published papers that propose side-channel attacks on HQC include the paper \emph{Multi-Value Plaintext Checking and Full-Decryption Oracle-Based Attacks on HQC from Offline Templates}~\cite{Dong_Guo_2025} and the paper \emph{Key Recovery from Side-Channel Power Analysis Attacks on Non-SIMD HQC Decryption}~\cite{10.1007/978-3-032-01901-1_3}.

The remainder of this paper is organized as follows: Section~\ref{sec:hqc} provides a brief description of HQC, while Section~\ref{sec:target_impl} describes the target implementation. In Section~\ref{sec:setup} and~\ref{sec:approach}, we present the setup and approach of our attack. The evaluation of the attack is discussed in Section~\ref{sec:evaluation}. In Section~\ref{sec:countermeasures}, we introduce possible countermeasures, and in Section~\ref{sec:time}, we analyze their time complexity. The paper concludes with Section~\ref{sec:conclusion}, which summarizes the results of our work.

\section{\uppercase{HQC}}
\label{sec:hqc}

HQC~\cite{HQC} is an IND-CCA2 secure code-based Key Encapsulation Mechanism (KEM) scheme. The security of the scheme is based on the Quasi-Cyclic Syndrome Decoding (QCSD) problem. The HQC KEM is derived from the IND-CPA secure HQC Public Key Encryption (PKE) scheme by using the Fujisaki-Okamoto transformation~\cite{fujisaki_okamoto_1999}.

The HQC PKE consists of three algorithms -- encryption, decryption, and key generation. Since our attack specifically targets the decryption algorithm, we describe it together with key generation in more detail below.

\subsection{HQC Key Generation and Decryption}
\label{subsec:hqc_key_decrypt}

During the key generation algorithm, a public key and a private key are produced. The private key consists of two sparse binary polynomials $(x,y)$, and the public key is given by the pair $(h,s)$, where
\[
    s = x + h \cdot y.
\]

During decryption, an intermediate value is computed by multiplying the ciphertext component~$u$ with the private key polynomial~$y$. Specifically, the decryption involves computing
\[
    \tilde{v} = v - u \cdot y,
\]
where $(u,v)$ is the ciphertext. The resulting vector~$\tilde{v}$ is then passed to the decoder to recover the plaintext message~$m$. The recovered message is then used to derive the shared secret key.

In this paper, we propose an attack that exploits side-channel leakage that occurs during the polynomial multiplication $u\cdot y$. More specifically, our attack focuses on recovering the value of the private key polynomial~$y$.

\section{\uppercase{Target Implementation}}
\label{sec:target_impl}

The authors of HQC submitted several different implementations. However, the target implementation of our attack is the \emph{Additional implementation} included in the NIST Round~4 submission~\cite{HQCsubmission}. We selected this implementation because it is included in both the PQClean~\cite{pqclean} and liboqs~\cite{liboqs} libraries, as noted in the introduction.

The target of our attack is a polynomial multiplication implemented using a recursive Karatsuba algorithm. Once the input operands in the recursive calls of the Karatsuba algorithm reach a size of 64 bits, a base-case 64-bit multiplication algorithm is used.

In the following sections, we analyze this base-case multiplication algorithm and propose a power analysis attack against its implementation. The attack recovers only a 64-bit portion (a single limb) of the private key. However, by repeating the attack on different limbs, an attacker can potentially recover the entire private key.

\subsection{Multiplication Algorithm}
\label{subsec:mul_alg}

The implementation of the multiplication algorithm is shown in Listing~\ref{list:base_mul}. The implementation is based on the mul1 algorithm presented in the paper Faster multiplication in GF(2)[x]~\cite{mul1}.

The implementation multiplies two 64-bit values, $a$ and $b$, using the window method. The result of this multiplication is stored in two variables, $l$ and $h$.

In the first step of the algorithm, a lookup table is created. This table has 16 entries and stores multiples of $b$. Before calculating and storing the multiples of $b$, the four highest bits are masked out. This is done in order to prevent $b$ from overflowing.

In the second step, the algorithm iterates over the bits of $a$, processing four bits during each iteration. These four bits are used to retrieve a value stored in the lookup table. This value is then added to $l$ and $h$ using XOR and shift operations.

To protect against cache timing attacks, all values in the lookup table are accessed during each iteration, but only the value stored at the index corresponding to the four processed bits is actually added to $l$ and $h$.

After step 2, the four masked out bits of $b$ are multiplied by $a$ and then added to the result. The result is then stored in the array $c$.

\begin{lstlisting}[style=Cstyle,caption={Function from \texttt{gf2x.c} from the \emph{Additional implementation} of HQC that performs multiplication of $a$ and $b$.},captionpos=b,label={list:base_mul}]
void base_mul(uint64_t *c, uint64_t a, uint64_t b) {

   uint64_t h = 0;
   uint64_t l = 0;
   uint64_t g;
   uint64_t u[16] = {0};
   uint64_t mask_tab[4] = {0};
   
   // Step 1
   u[0] = 0;
   u[1] = b & ((1UL << (64 - 4)) - 1UL);
   u[2] = u[1] << 1;
   u[3] = u[2] ^ u[1];
   u[4] = u[2] << 1;
   u[5] = u[4] ^ u[1];
   u[6] = u[3] << 1;
   u[7] = u[6] ^ u[1];
   u[8] = u[4] << 1;
   u[9] = u[8] ^ u[1];
   u[10] = u[5] << 1;
   u[11] = u[10] ^ u[1];
   u[12] = u[6] << 1;
   u[13] = u[12] ^ u[1];
   u[14] = u[7] << 1;
   u[15] = u[14] ^ u[1];

   // Step 2
   g=0;
   uint64_t tmp1 = a & 15;
   
   (*@\colorbox{yellow!30}{for(int i = 0; i < 16; i++) \{}@*)
     (*@\colorbox{yellow!30}{uint64\_t tmp2 = tmp1 - i;}@*)
     (*@\colorbox{yellow!30}{g \string^= (u[i] \& -(1 - ((tmp2 | -tmp2) >> 63)));}@*)
   (*@\colorbox{yellow!30}{\}}@*)
   
   l = g;
   h = 0;

   for (uint8_t i = 4; i < 64; i += 4) {
      g = 0;
      uint64_t tmp1 = (a >> i) & 15;
      
      (*@\colorbox{yellow!30}{for (int j = 0; j < 16; ++j) \{}@*)
         (*@\colorbox{yellow!30}{uint64\_t tmp2 = tmp1 - j;}@*)
         (*@\colorbox{yellow!30}{g \string^= (u[j] \& -(1 - ((tmp2 | -tmp2) >> 63)));}@*)
      (*@\colorbox{yellow!30}{\}}@*)

      l ^= g << i;
      h ^= g >> (64 - i);
   }

   // Step 3
   mask_tab [0] = - ((b >> 60) & 1);
   mask_tab [1] = - ((b >> 61) & 1);
   mask_tab [2] = - ((b >> 62) & 1);
   mask_tab [3] = - ((b >> 63) & 1);

   l ^= ((a << 60) & mask_tab[0]);
   h ^= ((a >> 4) & mask_tab[0]);

   l ^= ((a << 61) & mask_tab[1]);
   h ^= ((a >> 3) & mask_tab[1]);

   l ^= ((a << 62) & mask_tab[2]);
   h ^= ((a >> 2) & mask_tab[2]);

   l ^= ((a << 63) & mask_tab[3]);
   h ^= ((a >> 1) & mask_tab[3]);

   c[0] = l;
   c[1] = h;
}
\end{lstlisting}

\subsection{Implementation Analysis}
\label{subsec:mul_analysis}

As described, the multiplication algorithm consists of three main steps that multiply $a$ and $b$. In the target implementation of HQC, the value stored in $a$ serves as the private key. Therefore, the objective of our attack is to recover $a$ by analyzing power consumption.

To achieve this, we target the second step of the algorithm, which uses the value stored in $a$ to access the values in the lookup table.

To recover the value stored in $a$, we exploit the countermeasure designed to defend against cache timing attacks. This countermeasure works by iterating through the entire lookup table, but only using the value stored in the lookup table at the index equal to the value of the four bits of $a$. This vulnerable part occurs in the code twice -- for the first 4 bits and then for the rest. Both appearances are highlighted in Listing~\ref{list:base_mul} on lines (31-34) and (43-46).

By identifying, through power consumption analysis, which iteration causes a value to be stored in the temporary variable $g$, we can deduce the value of the four bits of $a$ currently processed. Repeating this process for all 16 sets of four bits allows us to fully recover the value of $a$.

\section{\uppercase{Attack Setup}}
\label{sec:setup}

To carry out the simple power analysis attack successfully, we need to measure the power consumption of the target device during the execution of the multiplication. For this purpose, we use the ChipWhisperer-Lite board, which is part of the open-source toolchain ChipWhisperer developed by the company NewAE Technology Inc. specifically for side-channel analysis.

\begin{figure}[!h]
  \centering
   {\epsfig{file = 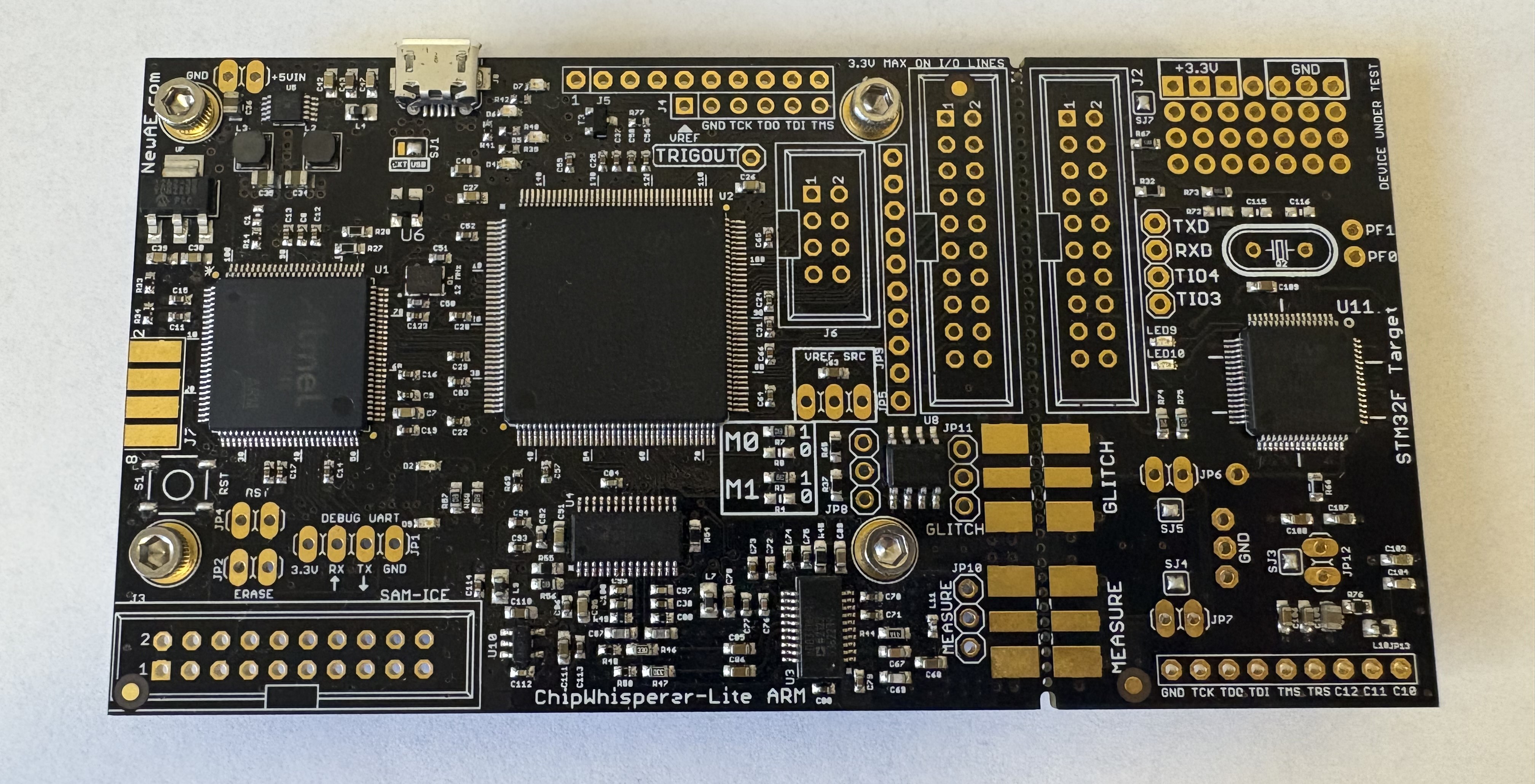, width = 7cm}}
  \caption{Figure of the ChipWhisperer-Lite board with a 32-bit STM32F303 target.}
  \label{fig:cwlite}
\end{figure}

The ChipWhisperer-Lite board, shown in Figure~\ref{fig:cwlite}, consists of two main components: the target board and the main board. The target is a 32-bit Arm Cortex-M4 based microcontroller (STM32F303), which can be programmed with the target implementation of the multiplication algorithm. The main board is responsible for measuring the power consumption of the target during execution.

To communicate with the ChipWhisperer-Lite, the board must be connected to a computer with a micro USB.

The ChipWhisperer-Lite supports various configuration options. In our work, we used the following configuration:
\begin{itemize}
    \item \textbf{Clock generator frequency:} \\
    \(\texttt{scope.clock.clkgen\_freq} \approx~7.38\,\text{MHz}\)
    
    \item \textbf{ADC sampling frequency} \\
    \(\texttt{scope.clock.adc\_freq} \approx~7.38\,\text{MHz}\)

    \item \textbf{Clock source for ADC:} \texttt{clkgen\_x1}
    
    \item \textbf{Number of samples per capture:} \\
    \(\texttt{scope.adc.samples} = 7500\)
    
    \item \textbf{Low-noise amplifier gain:} \\
    \(\texttt{scope.gain.db} \approx~24.84\,\text{dB}\)
\end{itemize}

For reproducibility of the attack, note that the vulnerable code was compiled with the default compiler optimization flags of ChipWhisperer version 5.6.1, including \texttt{-Os}. The power consumption traces and attack results presented in the following sections correspond to this build.

Different compilers or optimization flags may change the shape and timing of the traces. However, this does not prevent exploiting the same vulnerability on other builds, although adjustments to the attack may be necessary.

\section{\uppercase{Attack Approach}}
\label{sec:approach}

\begin{figure}[!h]
  \centering
   {\epsfig{file = 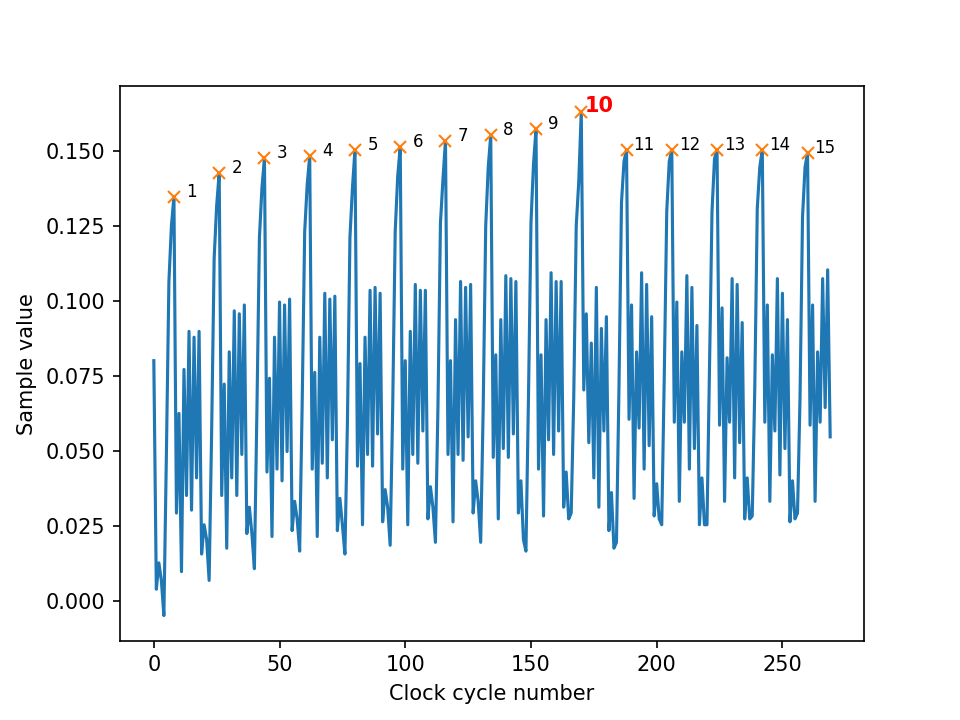, width = 7cm}}
  \caption{Power consumption trace used to recover the value of the first four bits of $a$, indicating a value of 10 based on the drop between peaks number 10 and 11.}
  \label{fig:spa1}
\end{figure}

\begin{figure}[!h]
  \centering
   {\epsfig{file = 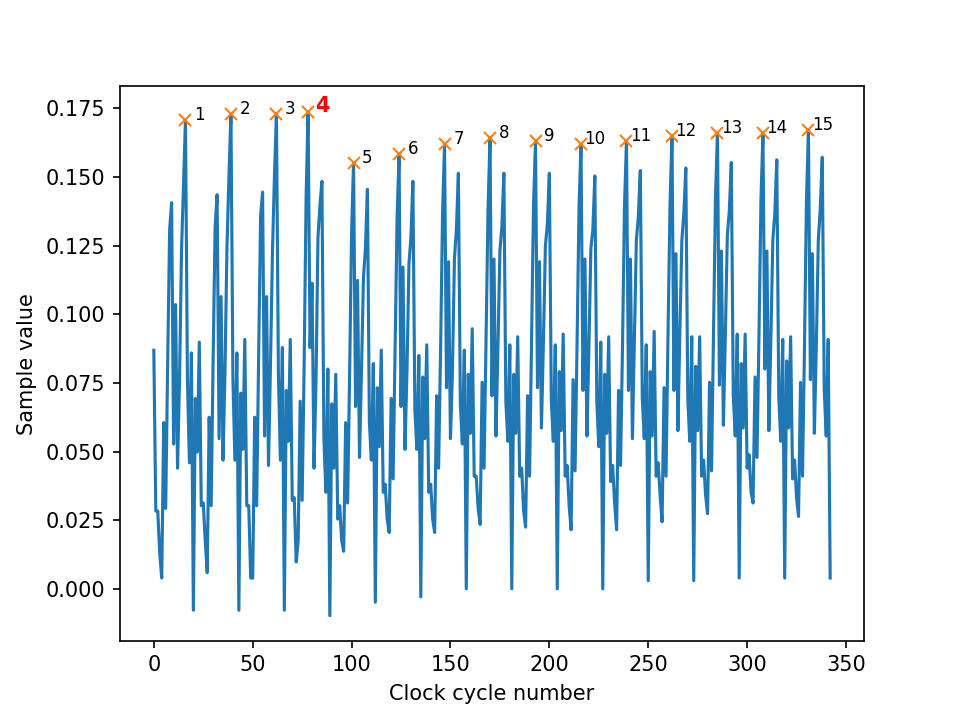, width = 7cm}}
  \caption{Power consumption trace used to recover the value of the second set of four bits of $a$, indicating a value of 4 based on the drop between peaks number 4 and 5.}
  \label{fig:spa2}
\end{figure}

To clearly illustrate the attack, we will demonstrate it by showing small segments of power consumption captured during multiplication performed with the first four bits of $a$ having a value of 10, and the second set of four bits having a value of 4. Specifically, Figure~\ref{fig:spa1} shows a trace of power consumption captured while the algorithm used the first four bits of $a$ to access the lookup table, and Figure~\ref{fig:spa2} shows a trace captured while the algorithm used the second set of four bits to access the lookup table.

To obtain these two figures, we first captured a single power consumption trace, then cropped it and identified peaks within the cropped trace using the \texttt{find\_peaks} function from the Python library \texttt{SciPy}.

After cropping the trace and identifying the peaks, we can observe that at a certain point in the traces, there is a noticeable drop between two neighboring peaks. By counting the number of peaks from the left up to this drop, we obtain the value that the corresponding four bits had during the execution of the multiplication.

It is important to note that the last peak before the drop is considerably higher than all other neighboring peaks. This characteristic is specifically used in our attack to differentiate between the four bits of $a$ that have a value of 0 or 15.

Another observation we made is that, apart from the first four bits of $a$, the values of the remaining 15 sets of four bits can be determined in a slightly different and easier way. As shown in Figure~\ref{fig:spa2}, each main peak is accompanied by a smaller secondary peak either on its left or right. By counting from the left and identifying the first main peak with its smaller secondary peak on the right, we obtain the value of the corresponding four bits of $a$.

\section{\uppercase{Attack Evaluation}}
\label{sec:evaluation}

To make evaluating the single-trace SPA attack as simple as possible, we automated the entire process by creating a Python script. The only human input required is the starting point and length of the power consumption segments. These inputs are used by the script to crop the power consumption trace and determine the values of all 16 sets of four bits of $a$, as explained in Section~\ref{sec:approach}.

To evaluate the attack, we performed it 10\,000 times with random values of $a$ and $b$. The overall success rate of the attack was 99.69\%, with 31 unsuccessful attempts. Notably, all failed attacks were due to errors in obtaining the first four bits of $a$, and none of the remaining 60 bits caused any failures.

\begin{figure*}[h]
  \centering
   {\epsfig{file = 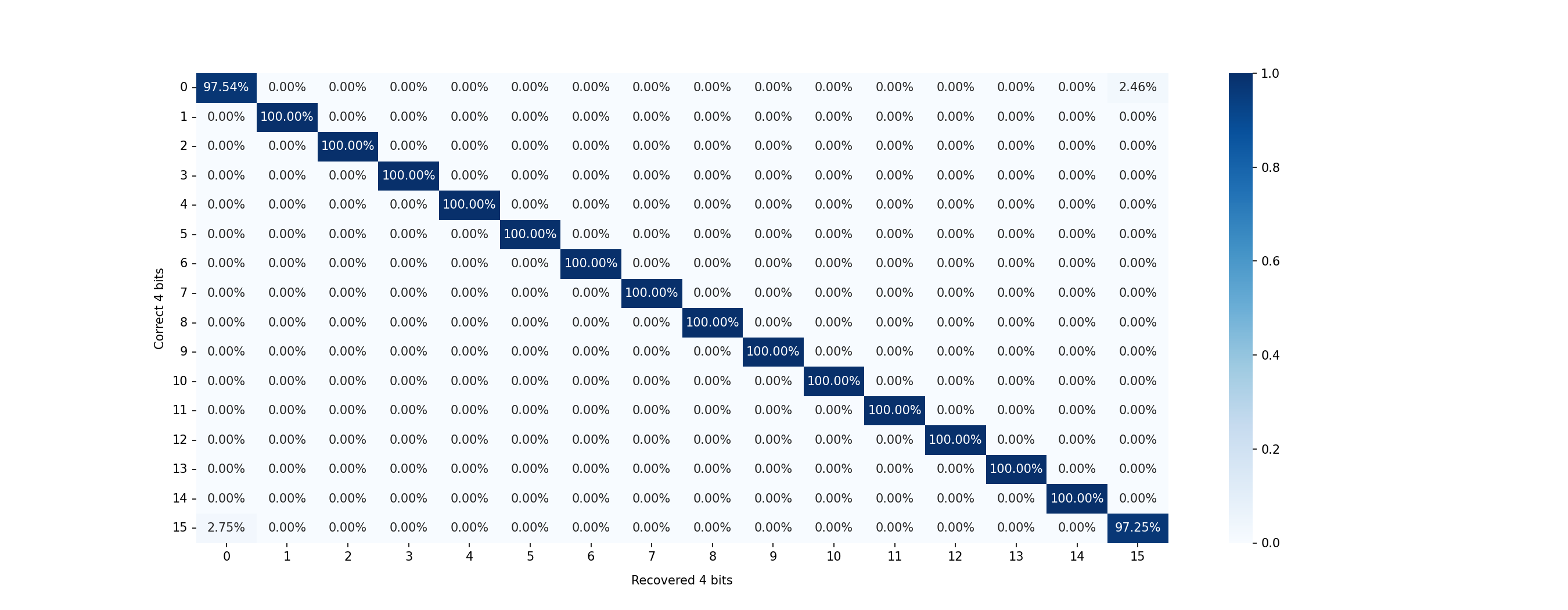, width = 18cm}}
  \caption{Normalized confusion matrix showing the actual correct values along with the values recovered by the single-trace SPA attack for the first 4 bits of $a$. The matrix is row-normalized and shows the distribution of the recovered values for each actual value.}
  \label{fig:cm4}
\end{figure*}

\begin{figure*}[h]
  \centering
   {\epsfig{file = 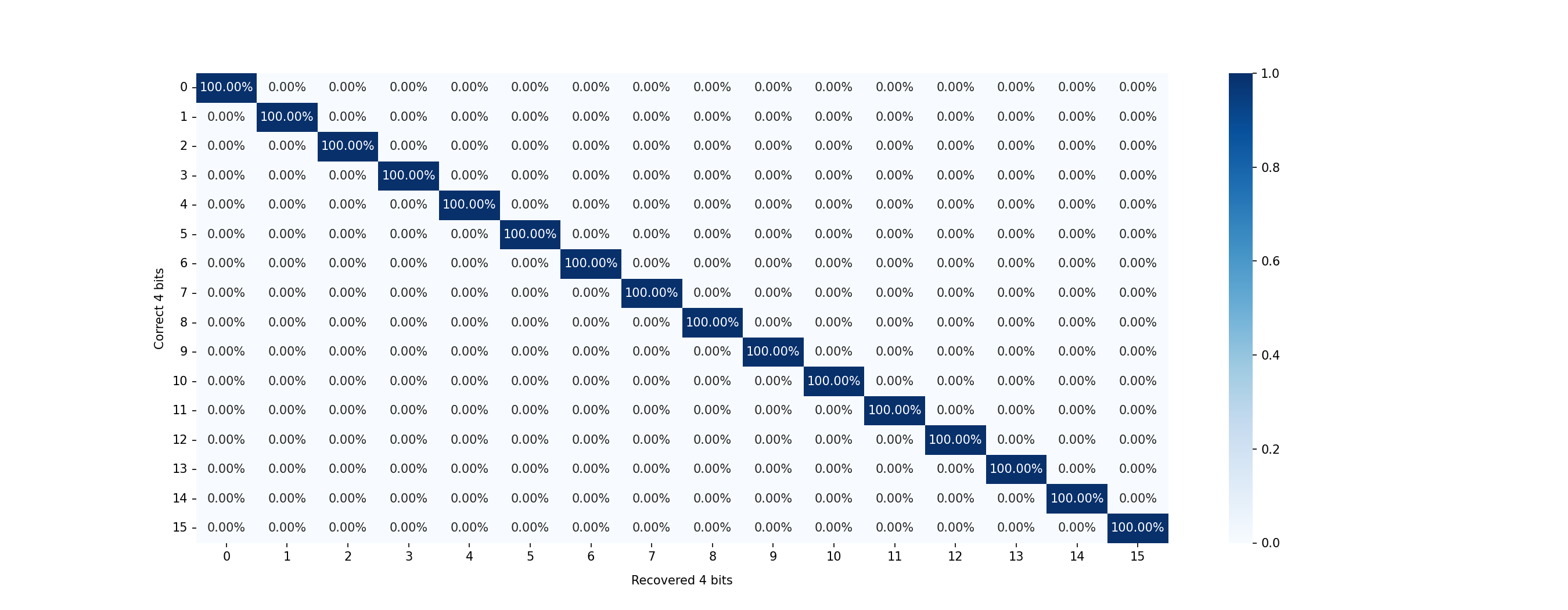, width = 18cm}}
  \caption{Normalized confusion matrix showing the actual correct values along with the values recovered by the single-trace SPA attack for the last 60 bits of $a$. For this portion of the private key, all bits were recovered with 100\% accuracy. The matrix is row-normalized and shows the distribution of the recovered values for each actual value.}
  \label{fig:cm60}
\end{figure*}

Figures~\ref{fig:cm4} and \ref{fig:cm60} show the confusion matrices corresponding to these results. These matrices illustrate the actual correct bit values along with the values recovered by the attack. It shows that all errors originate from the first four bits of $a$, and more specifically, only when the correct values were 0 or 15.

The results of our attack evaluation demonstrate the feasibility of recovering a 64-bit portion of the private key from a single power trace. By successfully repeating this attack for all calls of the multiplication algorithm that occur during polynomial multiplication of the ciphertext and the private key, an attacker can recover the entire private key.

\clearpage

\section{\uppercase{Countermeasures}}
\label{sec:countermeasures}

The attack presented in this work is possible due to the countermeasure implemented against cache timing attacks. This countermeasure works by accessing all values stored in a lookup table whenever a single value is needed. While this approach effectively protects against cache timing attacks, it also introduces a new vulnerability that attackers can potentially exploit, as demonstrated in this work.

Another issue with the countermeasure against cache timing attacks is that, although it successfully prevents these attacks, it also causes the algorithm to lose the performance advantage that motivated its implementation. As a result, the intended speed improvement is not achieved. The time complexity of different implementations of the multiplication is discussed in more detail in Section~\ref{sec:time}.

\begin{lstlisting}[style=Cstyle,caption={Implementation of multiplication algorithm that does not use a lookup table.},captionpos=b,label={list:base_mul2}]
void base_mul2(uint64_t *c, uint64_t a, uint64_t b) {
    uint64_t mask = - (a & 1);
    uint64_t l = b & mask;
    uint64_t h = 0;

    for (uint64_t i = 1; i < 64; i++) {
        uint64_t mask = - ((a >> i) & 1);
        l ^= (b << i) & mask;
        h ^= (b >> (64 - i)) & mask;
    }
    
    c[0] = l;
    c[1] = h;
}
\end{lstlisting}

The first approach to defending against the attack presented in this work is to reduce the size of the lookup table to one. In other words, instead of using a lookup table that stores multiples of $b$, we can simply use $b$ itself. By not using the lookup table, the need to implement a countermeasure against cache timing attacks is removed, along with the first and third steps of the algorithm. The disadvantage of this approach is that it eliminates the theoretical speed improvement offered by the multiplication algorithm that uses a lookup table.

The implementation of multiplication algorithm that does not use a lookup table is shown in Listing~\ref{list:base_mul2}.

Another possible approach to defending against the single-trace SPA attack presented in this work is to remove the original countermeasure against cache timing attacks and instead implement alternative countermeasures.

The Listing~\ref{list:base_mul3} shows an implementation of the multiplication algorithm without the countermeasure against cache timing attacks. The listing omits the initialization as well as the first and third steps, since they are identical to those in Listing~\ref{list:base_mul}.

\begin{lstlisting}[style=Cstyle,caption={Implementation of the multiplication algorithm without the countermeasure against cache timing attacks. The initialization as well as the first and third steps are omitted, since they are identical to those in Listing~1.},captionpos=b,label={list:base_mul3}]
void base_mul3(uint64_t *c, uint64_t a, uint64_t b) {

    // ... initialization and Step 1 ...

    // Step 2
    g=0;
    uint64_t tmp = a & 15;
    g ^= u[tmp];
    l = g;
    h = 0;

    for (uint8_t i = 4; i < 64; i += 4) {
        g = 0;
        uint64_t tmp = (a >> i) & 15;
        g ^= u[tmp];
        l ^= g << i;
        h ^= g >> (64 - i);
    }

    // ... Step 3 ...

    c[0] = l;
    c[1] = h;
}
\end{lstlisting}

\subsection{Masking}
\label{subsec:countermeasures_mask}

While the implementation shown in Listing~\ref{list:base_mul3} is not vulnerable to the SPA attack presented in this work, it may still be vulnerable to cache timing attacks, as well as other power analysis attacks.

A possible approach to protecting against these threats is the classical masking technique~\cite{powerAnalysisCards}, more specifically Boolean masking. With this approach, the value of $a$ is first masked using a randomly generated mask in the following way:
\begin{align*}
    aMasked = a \oplus mask
\end{align*}

The masked value of $a$ is then multiplied by $b$:
\begin{align*}
    c &= aMasked \times b = (a \oplus mask) \times b \\
    c &= (a \times b) \oplus (mask \times b)
\end{align*}

To remove the mask from the result of the multiplication, the mask must also be multiplied by $b$. The result of this multiplication can then be used to obtain the correct product of $a$ and $b$:
\begin{align*}
    c &= c \oplus (mask \times b) \\
    c &= (a \times b) \oplus (mask \times b) \oplus (mask \times b) = a \times b
\end{align*}

The Listing~\ref{list:masking} shows an example implementation of the masking countermeasure.

\begin{lstlisting}[style=Cstyle,caption={Implementation of masking designed to protect the value of $a$ against cache timing attacks and power analysis attacks.},captionpos=b,label={list:masking}]
    a = a ^ mask;
    uint64_t c[2];
    base_mul(c,a,b);

    uint64_t unmask[2];
    base_mul(unmask,mask,b);
    
    c[0] = c[0] ^ unmask[0];
    c[1] = c[1] ^ unmask[1];
\end{lstlisting}

It is important to note that, while this masking technique is effective, an attacker may still be able to carry out successful attacks on the two performed multiplications and potentially recover the masked value of $a$ and the mask value. With these two values, the attacker can then obtain the true value of $a$.

\section{\uppercase{Multiplication Algorithm Complexity}}
\label{sec:time}

To analyze the new implementations in more detail, we measured the CPU time required to execute the functions \texttt{base\_mul} (original implementation), \texttt{base\_mul2} (without lookup table), and \texttt{base\_mul3} (without countermeasure against cache timing attacks) 10\,000\,000 times. Table~\ref{tab:execution_times} shows the corresponding measured values.

The execution times were measured on a system with the following specifications: Intel Core i5-8350U CPU @ 1.70 GHz, 8 GB RAM, Windows 11. All implementations were compiled using \texttt{gcc} version 14.2.0 with the \texttt{-Os} optimization flag.

\begin{table}[h]
\caption{Execution times (in seconds) for 10\,000\,000 calls of the functions \texttt{base\_mul}, \texttt{base\_mul2}, and \texttt{base\_mul3}.}\label{tab:execution_times} \centering
\begin{tabular}{c|c|c}
  \texttt{base\_mul} & \texttt{base\_mul2} & \texttt{base\_mul3} \\
  \hline
   3.244188\,s & 1.434326\,s & 0.472569\,s \\
\end{tabular}
\end{table}

As shown in Table~\ref{tab:execution_times}, the function \texttt{base\_mul2} is approximately $2.3\times$ faster than \texttt{base\_mul}, while \texttt{base\_mul3} is approximately $6.9\times$ faster.

This shows that \texttt{base\_mul2} and \texttt{base\_mul3} are not only protected against the attack presented in this work, but also faster than \texttt{base\_mul}. Although these functions may still be vulnerable to other power analysis attacks (or, in the case of \texttt{base\_mul3}, to cache timing attacks), they can be secured using the masking countermeasure introduced in section~\ref{subsec:countermeasures_mask}, while still maintaining a performance advantage over \texttt{base\_mul}.

\section{\uppercase{Conclusion}}
\label{sec:conclusion}

In this paper, we proposed a new single-trace simple power analysis (SPA) attack against the post-quantum cryptosystem HQC. We demonstrated how a significant portion of the private key can be recovered by analyzing power consumption during execution of the multiplication function \texttt{base\_mul}. This function serves as the base case for the Karatsuba algorithm in the \emph{Additional implementation} of HQC included in the NIST Round~4 submission~\cite{HQCsubmission}. The same implementation is also included in both the PQClean and liboqs libraries increasing the relevance of our findings due to their popularity.

To perform and evaluate the proposed SPA attack, we used the ChipWhisperer platform, specifically the ChipWhisperer-Lite. Using a single power consumption trace per attack attempt, we performed 10\,000 attacks and achieved a 99.69\% success rate.

We also implemented two new multiplication functions, \texttt{base\_mul2} and \texttt{base\_mul3}, which are protected against the proposed SPA attack. After implementing these functions, we evaluated their performance and found that they are not only resistant to our attack but also offer a significant performance advantage. In our experiments, \texttt{base\_mul2} was approximately $2.3\times$ faster than \texttt{base\_mul}, while \texttt{base\_mul3} was approximately $6.9\times$ faster.

In future work, we plan to further analyze other implementations of HQC and assess their resistance against side-channel attacks.

\section*{\uppercase{Acknowledgements}}

This work was supported by the Student Summer Research Program 2025 of FIT CTU in Prague and by the Grant Agency of the Czech Technical University in Prague, grant No. SGS23/211/OHK3/3T/18 funded by the MEYS of the Czech Republic.

\bibliographystyle{apalike}
{\small
\bibliography{SPAattackOnHQC}}

\end{document}